\documentclass[invmat]{svjour}
\usepackage{color}
\usepackage{amsmath}
\usepackage{verbatim}
\usepackage{graphicx}
\usepackage{psfrag}
\usepackage{mathtools}

\def\x{{\rm x}}
\def\k{{D}}

\def\bfR{{\rm R}}

\def\f{{\bf J}}
\def\div{{\rm div}}

\def\DT{{\cal D}_T}
\def\ST{{\cal S}_T}
\begin{document}
\title{Einstein's random walk and thermal diffusion}
\author{
Yong-Jung Kim
}

\institute{Department of Mathematical Sciences, KAIST \\
 291 Daehak-ro, Yuseong-gu, Daejeon 305-701, Korea}
\date{\today}
\maketitle
\begin{abstract}
Thermal diffusion has been studied for over 150 years. Despite of the long history and the increasing importance of the phenomenon, the physics of thermal diffusion remains poorly understood. In this paper Ludwig's thermal diffusion is explained using Einstein's random walk. The only new structure added is the spatial heterogeneity of the random walk to reflect the temperature gradient of thermal diffusion. Hence, the walk length and the  walk speed are location dependent functions in this paper. Then, a mathematical understanding of such a random walk gives the foundation of the thermal diffusion as clearly as the original homogeneous case of Einstein.
\end{abstract}

\section*{Introduction}

``Molecules drift along temperature gradients, an effect called thermophoresis, the Soret effect, or thermodiffusion. In liquids, its theoretical foundation is the subject of a long-standing debate" (quoted from \cite{Duhr2006}).  The purpose of this article is to show that the theoretical foundation of thermal diffusion is still Einstein's random walk and to derive the non-isothermal diffusion law that combines the normal diffusion theory and the thermal diffusion one.

In 1855, Adolf Fick \cite{Fick1855} found that the diffusion of salt concentration has an analogy to the heat conduction and proposed a diffusion flux,
\begin{equation}\label{FickLaw}
\f=-\k \nabla u,\quad \k\ge0,
\end{equation}
where this diffusion flux is simply an analogy of Fourier's law of heat conduction. Here, $u$ is the particle concentration in space dimensions $n\ge1$, $\nabla u=\big({\partial u\over\partial x_1},\cdots,{\partial u\over\partial x_n}\big)$ is the gradient vector field, and, according to Fick, the diffusivity $\k$ is a constant depending on the nature of the substances. Fifty years later, in 1905, Albert Einstein \cite{Einstein1905} explained the physics of Fick's phenomenological diffusion theory in terms of Brownian motion or random walks. He showed that the diffusivity $\k$ is given by
\begin{equation}\label{ES1}
\k={1\over2n}{\langle x^2\rangle\over t},
\end{equation}
which is now called the Einstein-Smoluchowski relation. Here, the numerator $\langle x^2\rangle$ denotes the variance of the location probability of a Brownian particle that started the origin and traveled for a time length $t>0$. This relation holds for any given time length $t>0$. If it is taken as small as the mean collision time of Brownian motion, say $\Delta t$, then the relation can be written as
$$
\k={1\over2n}{|\Delta x|^2\over \Delta t},
$$
where $\Delta x$ is the mean free path of Brownian particles in the root mean square sense. Einstein's random walk  and his theoretical foundation of diffusion in the molecular level kick-started a revolution in statistical physics and in many other fields that the randomness plays a key role (see \cite{Haw2005}).

In the meanwhile, Carl Ludwig \cite{Ludwig1856} found in 1856 that, if a temperature gradient $\nabla T$ is applied across a uniformly distributed salt solution, the salt particles move toward colder regions and a concentration gradient is formed. This phenomenon is called a thermodiffusion or thermophoresis which cannot be explained by Fick's diffusion law (\ref{FickLaw}). The importance of thermal diffusion in various phenomena and possible applications in emerging nano- and bio-technologies or isotope fractionation can be found from  \cite{Braun2004,Duhr2006,Huang2010,Putnam2007} and references therein. There have been enormous amount of researches related to this thermal diffusion phenomenon and readers are referred to experimental and modeling review papers \cite{Eslamian2009,Harstad2009,Srinivasan2011} and references therein. However, the consensus in the literature is that there is no comprehensive or generic thermal diffusion models such as Einstein's molecular level explanation for the homogeneous case.

Let us return to Ludwig's observation and explain a thermophoresis model. Since the salt particles move to colder regions, the corresponding flux is phenomenologically modeled by $-u\DT\nabla T$ and is called a thermal force, where the scaling coefficient $\DT $ is called thermal diffusivity. The whole flux for this thermal diffusion model is given by combining it to the normal concentration diffusion, i.e.,
\begin{equation}\label{ThermalDiffusion}
\f=-D\nabla u-u\DT \nabla T.
\end{equation}
Then the steady state is obtained when the concentration diffusion and the thermal diffusion are balanced, i.e.,
$$
0=-D\nabla u-u\DT \nabla T\mbox{~~~~or~~~}{1\over u}\nabla u=-\ST \nabla T,
$$
where $\ST :=\DT /D$ is called the Soret coefficient. However, this thermal force is only a phenomenological explanation and the source of such a force has never been verified. In fact, ``the thermal diffusion is the only hydrodynamic transport mechanism that lacks a simple physical explanation" and ``there is so far no molecular understanding of thermodiffusion in liquids" (see \cite{Kincaid1994,Wiegand2004} for more discussions). We will also see that the thermal diffusion theory for gaseous states fails.

\section*{Diffusion by non-uniform random walk}

In this article it is shown that the source of the mysterious thermal force is simply the randomness of Brownian motion. The difference from the normal diffusion is that, under a thermal gradient, Brownian displacement is not spatially homogeneous anymore and it is this heterogeneity that produces such a thermal flux.

\begin{figure}[ht]
\centering
    \psfrag{w-1}{$U^{i-1}$}    \psfrag{w+1}{$U^{i+1}$}    \psfrag{w}{$U^i$}    \psfrag{xi-1}{$x^{i-1}$}    \psfrag{xi+1}{$x^{i+1}$}    \psfrag{xi}{$x^i$}    \psfrag{delta x-1}{\hskip 1mm$\Delta x\big|_{x^{i-1/2}}$}    \psfrag{delta x+1}{\hskip 1mm$\Delta x\big|_{x^{i+1/2}}$}    \psfrag{alpha w-1}{\hskip -2mm ${U^{i-1}\over2}\ \  {U^{i-1}\over2}$}    \psfrag{alpha w}{\hskip -4mm ${U^{i}\over2}\qquad{U^{i}\over2}$}    \psfrag{alpha w+1}{\hskip -2mm ${U^{i+1}\over2}\ \  {U^{i+1}\over2}$}    \psfrag{delta t-1}{}    \psfrag{delta t+1}{}
  \includegraphics[width=110mm]{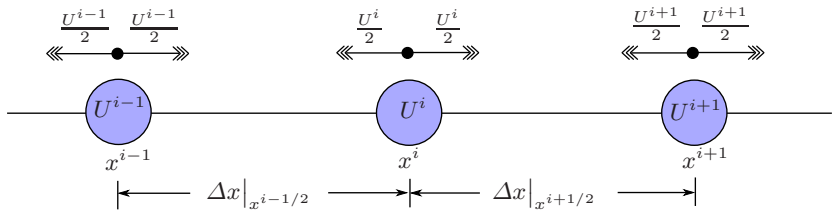}
\caption{Diagram of a random walk system}
\label{fig1}
\end{figure}

Let $x^i$ be grid points and $x^{i+1/2}:=(x^i+x^{i+1})/2$ be the midpoint between two adjacent grid points (see Figure \ref{fig1}). Let $U(x^i)$ be the number of particles at $x^i$, which are evenly distributed in the interval $(x^{i-1/2},x^{i+1/2})$. Each particle in the interval jumps randomly to one of two adjacent ones. Define the walk length at $x^{i+1/2}$ by $\Delta x\big|_{x^{i+1/2}} :=x^{i+1}-x^i$ and at $x^i$ by $\Delta x\big|_{x^{i}}:=x^{i+1/2}-x^{i-1/2}$. The traveling time at $x^i$ is denoted by $\Delta t\big|_{x^{i}}$, which is the length of time needed for a  particle to jump from $x^{i-1/2}$ to $x^{i+1/2}$ or vice versa. Similarly, the traveling time at $x_{i+1/2}$ is denoted by $\Delta t\big|_{x^{i+1/2}}$.

The half of the particles at the grid point $x^i$, evenly distributed in the interval $(x^{i-1/2},x^{i+1/2})$, will cross a midpoint $x^{i+1/2}$ during the traveling time $\Delta t\big|_{x^{i}}$. Hence, the flux of particles that crosses the midpoint $x^{i+1/2}$ from left to right is ${U\over2\Delta t}\big|_{x=x^{i}}$. Similarly, the flux from right to left is ${U\over2\Delta t}\big|_{x=x^{i+1}}$. Notice that the particle density is $u={U\over\Delta x}$ and hence the net flux across $x=x^{i+1/2}$ is
\begin{eqnarray*}
\f(x^{i+{1\over2}})&&={\Delta x\over2\Delta t}u\Big|_{x^{i}}-{\Delta x \over2\Delta t}u\Big|_{x^{i+1}}\\
&&=-{\Delta x\big|_{x^{i+{1\over2}}}\over2}\left(\frac{{\Delta x\over\Delta t}u\Big|_{x^{i}}-{\Delta x\over\Delta t}u\Big|_{x^{i+1}}} {~x^i~~~~-~~x^{i+1}~~}\right)\\
&&\cong\left.-{\Delta x\over2}
{\partial\over\partial x}\Big({\Delta x\over\Delta t}u\Big)\right|_{x=x^{i+{1\over2}}}.
\end{eqnarray*}
Notice that the $\Delta x$ in front of the parenthesis is to approximate the flux using a gradient and the ${\Delta x\over \Delta t}$ inside of it is to measure the flux itself. Hence they should be clearly distinguished. If the Brownian motion or the random walk is in a homogeneous environment, we may assume the walk length (or the mean free path) $\Delta x$ and the traveling time (or the mean collision time) $\Delta t$ are constant and hence ${\Delta x\over\Delta t}$ inside of the parenthesis can be taken out. However, if the temperature is not spatially constant, they depend on the space variable and should stay inside of it.

In conclusion, the diffusion flux for a non-isothermal case in $n$ space dimensions is given by
\begin{equation}\label{RW2}
\f= -{D\over S}\nabla\big(S u\big),\quad D:={|\Delta x|^2\over2n\Delta t},\quad S:={\Delta x\over\Delta t},
\end{equation}
where $S$ is the walk speed or the instantaneous velocity of a Brownian particle. Therefore, the corresponding non-isothermal diffusion equation is
\begin{equation}\label{non-iso-thermal}
u_t=\div\Big({D\over S}\nabla\big(S u\big)\Big),
\end{equation}
which is the diffusion model of this paper. In the model the role of two coefficients ${D\over S}$ and $S$ should be clearly distinguished. The $S$ inside of the gradient operator decides the steady state. For example, if $Su$ becomes constant, then the diffusion flux becomes zero and hence the steady state is inversely proportional to the walk speed $S$. On the other hand, the other coefficient ${D\over S}$ controls the speed to reach to this steady state.

It is not that surprising that the concentration density of steady state should be inversely proportional to the particle speed. In fact, there have been following kinds of speculations.  The length of trace of a freely moving Brownian particle in a region of the unit volume should be independent of its speed. Hence, the probability for the Brownian particle to stay in a region of unit volume is expected to be inversely proportional to the speed since so is the amount of time to stay in the region. The derivation above actually confirms such a speculation using a non-uniform random walk system.

\section*{Monte Carlo simulations}

Now we show that the new diffusion law (\ref{RW2}) or (\ref{non-iso-thermal}) explains random walk phenomena correctly by comparing the steady state given by the diffusion law to a Monte Carlo simulation.
\begin{figure}[ht]
\centering
\begin{minipage}[t]{50mm}
 \centering
 \includegraphics[width=50mm]{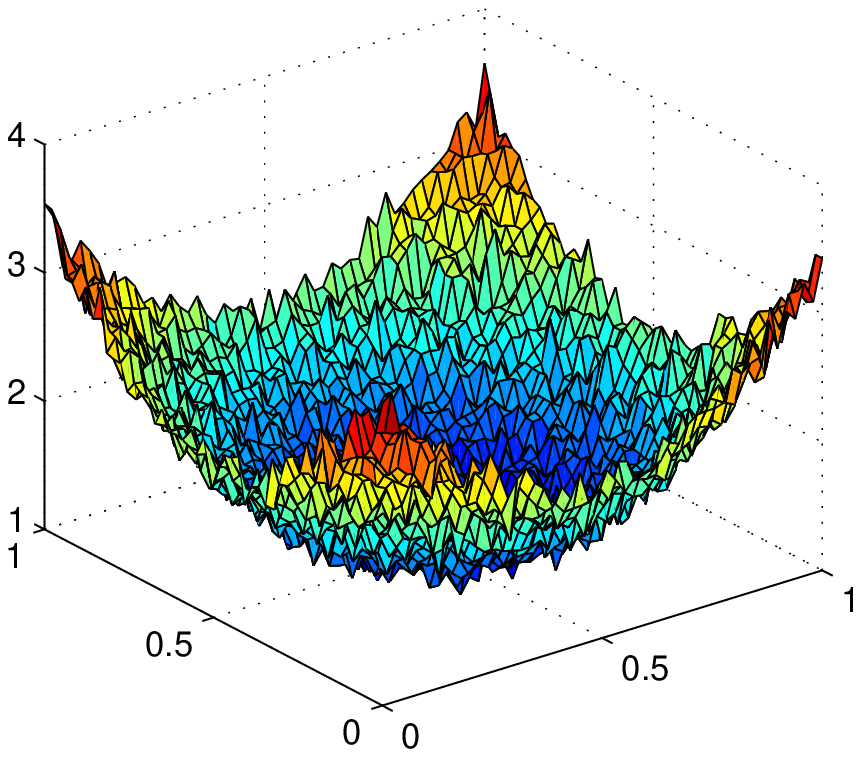}

(a) 1,000,000 particles.
\end{minipage}
\begin{minipage}[t]{50mm}
\centering
 \includegraphics[width=50mm]{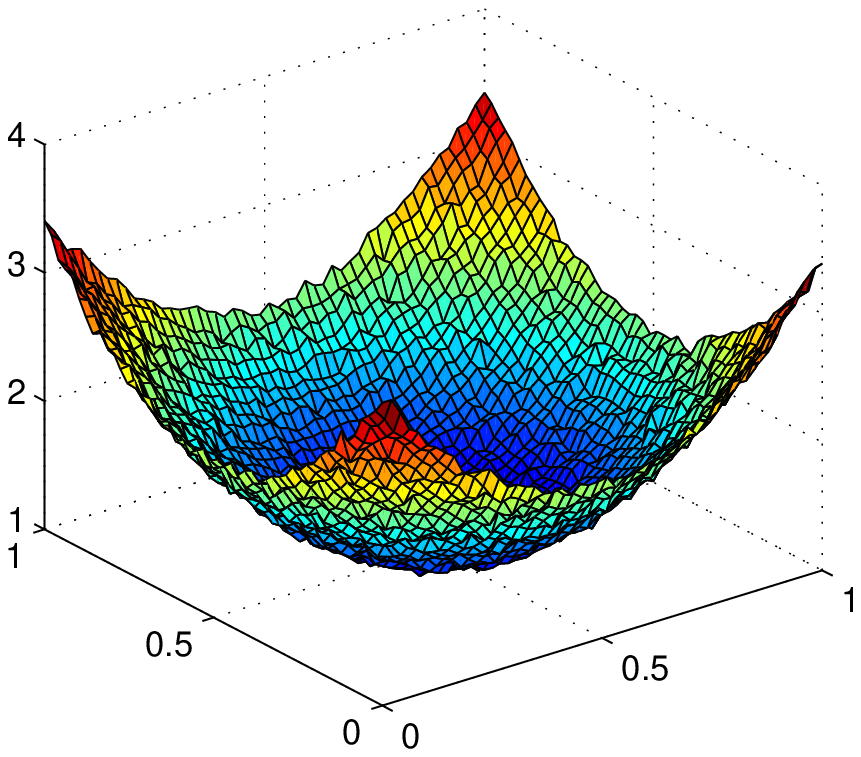}
(b) 10,000,000 particles.
\end{minipage}
\begin{minipage}[t]{50mm}
 \centering
 \includegraphics[width=50mm]{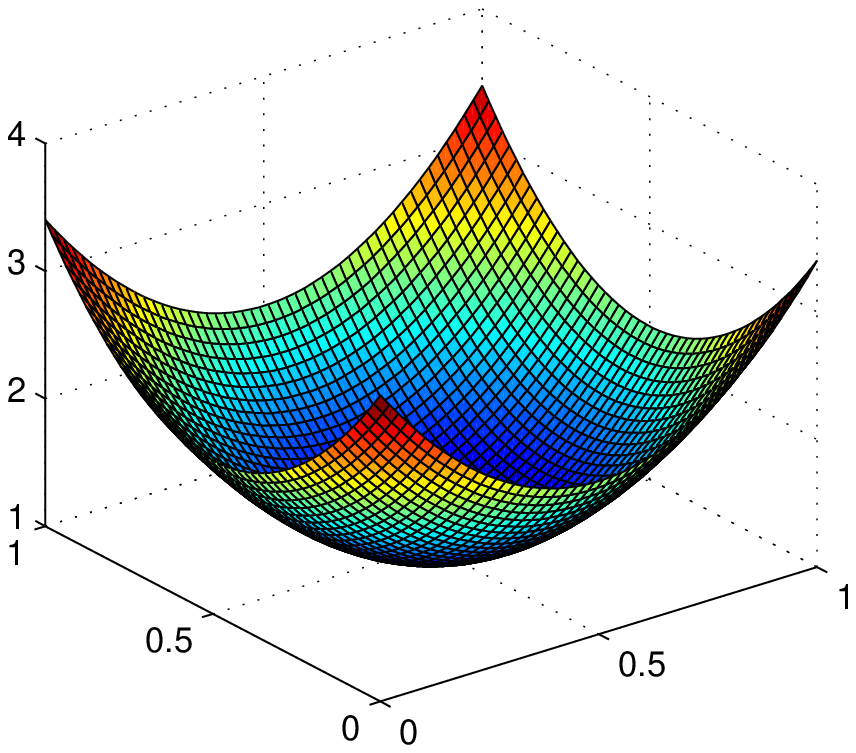}

(c) theoretical steady state
\end{minipage}
\begin{minipage}[t]{50mm}
\centering
 \includegraphics[width=50mm]{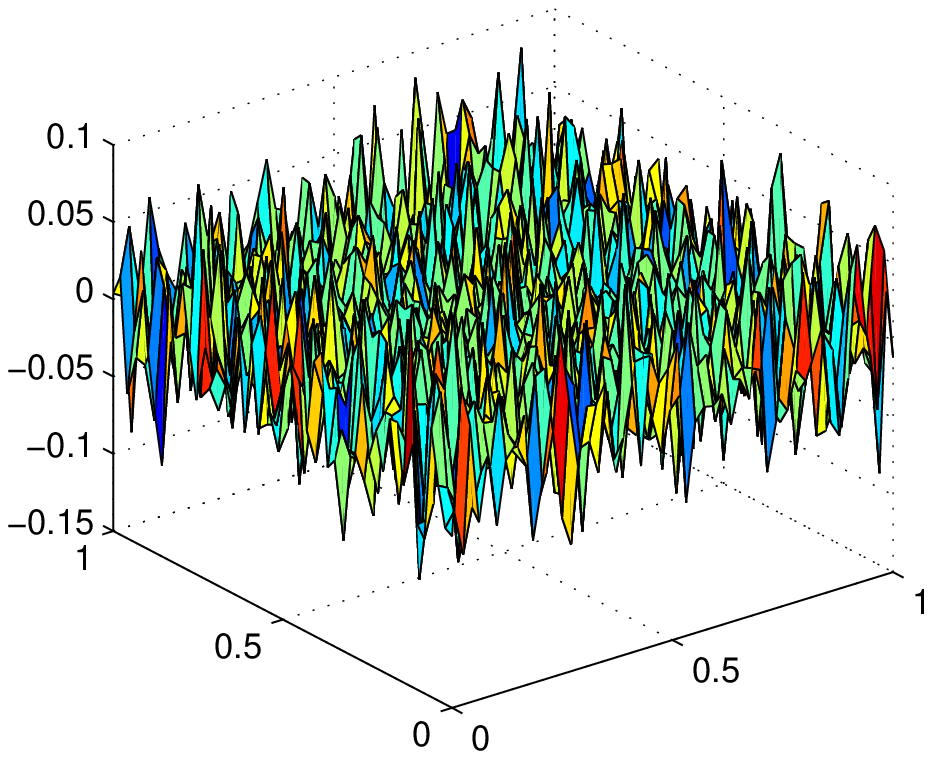}
(d) difference between (b) and (c).
\end{minipage}
\caption{Particle density distribution of a Monte Carlo simulation of a gridless non-uniform random walk in $\bfR^2$. The walk length is given by the function in (\ref{MCS3}). The domain was divided into $50\times50$ smaller regions and the number of particles in each subregions were counted. The figures are given after a normalization.} \label{fig2}
\end{figure}
We consider a random walk system in two space dimensions with non-constant $\Delta t$ and $\Delta x$. The domain is the unit square $\Omega=\{\x:=(x,y): 0<x,y<1\}\subset\bfR^2$. The boundary condition is periodic, i.e., $u(0,y)=u(1,y)$ and $u(x,0)=u(x,1)$. The walk length and the jumping time length are
\begin{equation}\label{MCS3}
\Delta x=0.02\times(0.2+|\x-0.5|^2),\quad \Delta t=0.02\times(0.2+|\x-0.5|^2)^2.
\end{equation}
In this case the diffusivity is constant but the speed $S$ isn't, i.e.,
\begin{equation}\label{DS2}
D=0.005,\qquad S=(0.2+|\x-0.5|^2)^{-1}.
\end{equation}
Therefore, the steady state of the non-isothermal diffusion equation (\ref{non-iso-thermal}) is proportional to $S^{-1}=0.2+|\x-0.5|^2$. The figure of this steady state is given in Figure \ref{fig2}(c) after a normalization.

In Figures \ref{fig2}(a) and \ref{fig2}(b), two Monte Carlo simulations are given using 1,000,000 and 10,000,000 particles respectively. In the simulation, each particles walk for time period $t>0$ with $2\sqrt{Dt\,}>10$, which is long enough to compare it to the steady state. One can observe that, as the number of particles are increased, the  particle density distribution converges to the steady state. In Figure \ref{fig2}(d), the difference between the theoretical steady state Figure \ref{fig2}(c) and the Monte Carlo simulation in Figure \ref{fig2}(b) is given. This figure shows that the distribution of the difference is uniform which confirms that the steady state in Figure \ref{fig2}(c) is the correct limit as $t\to\infty$.

The Monte Carlo simulation clearly shows that the steady state of the random walk is not constant even though the diffusivity is constant. This simple experiment implies a rather surprising consequence that it is not the diffusivity that decides the steady state of heterogeneous thermal diffusion. Diffusion models have been developed in terms of diffusivity (see, e.g., \cite{Bringuier2009,Chapman1928,MR0258399,Milligen2005}). It is also true that a lot of the experiments and theory are developed in terms of heterogeneous diffusivity. However, considering a heterogeneous diffusivity is not a correct approach to study a diffusion phenomenon in a heterogeneous environment.

\section*{Thermodynamics}

Einstein's idea of connecting random walks to thermodynamics and J. Perrin's \cite{Perrin1909} experimental proof for the existence of atoms made a revolutionary change in statistical physics and other related fields. We follow his idea and connect the dynamics of our non-uniform random walks to thermodynamics and compute the thermal diffusivity $\DT$ and the Soret coefficient $\ST$.

The walk speed $S={\Delta x\over\Delta t}$ of the random walk corresponds to the speed of Brownian particles which is a function of temperature. Therefore, if there is a temperature gradient, the location dependency of the walk speed is obtained through the temperature, i.e., $S=S(T)$. Then, the flux is written by
\begin{equation}\label{RW3}
\f= -{D\over S}\nabla\big(S u\big) = -\k\nabla u-u{D\over S}{dS\over dT}\nabla T.
\end{equation}
The thermal diffusion coefficient $\DT $ in (\ref{ThermalDiffusion}) has now a molecular description, which is given by
\begin{equation}\label{TDC}
\DT ={D\over S}{dS\over dT},
\end{equation}
and the Soret coefficient $\ST$ became
\begin{equation}\label{SC}
\ST ={1\over S}{d S\over d T}={d\over dT}\ln(S).
\end{equation}
Notice that we don't need to know the actual particle speed $S$ of Brownian particles to find the thermal diffusivity $\DT $ and the Soret coefficient $\ST$. For example, if $\bar S=cS$ is given without knowing the constant $c>0$, then $\DT$ and $\ST$ are simply given by the same relation, i.e.,
$$
{d\over dT}\ln(S)={d\over dT}\ln(cS)={d\over dT}\ln(\bar S),\quad
{D\over S}{dS\over dT}={D\over cS}{d(cS)\over dT}={D\over\bar S}{d\bar S\over dT}.
$$
Therefore, we may forget about the scaling coefficients in determining the thermal diffusivity $\DT$ and Soret coefficient $\ST$.

Let $M>0$ be the mass of a Brownian particle, $T>0$ be the temperature and $k_B$ be the Boltzman coefficient. Then, according to Einstein's equipartition theorem \cite{Einstein1907}, the particle speed $v$ satisfies
$$
{1\over2}Mv^2={3\over2}k_BT.
$$
The temperature $T$ is defined by the mean of the kinetic energy of particles and hence the velocity $v$ should be understood as the average velocity in root mean square sense and hence it exactly corresponds to the walk speed $S$ of a random walk system. Therefore, the speed $S$ is given by
\begin{equation}\label{Speed}
S\equiv v=\sqrt{k_BT/M}=c\sqrt{T},\quad c=\sqrt{k_B/M}.
\end{equation}

Measuring the actual speed of Brownian particles has a significance not only for an application aspect but also for theory itself. Einstein mentioned that its measurement will confirm the equipartition theorem. However, he was doubt if it can be really measured. Recently, T. Li {\it et al.} \cite{Velocity} actually measured it and confirmed the equipartition theorem for Brownian particles in gaseous state. These technological improvements may enable us to measure the instantaneous Brownian particle speed even in liquid state some day.

However, since the computation of the thermal diffusivity $\DT$ and the Soret coefficient $\ST$ is independent of the scale of $S$, we may compute the thermal diffusivity without knowing the instantaneous speed of Brownian particles. We will drop the constant $c$ above and simply set
$$S=\sqrt{T}$$
for the sake of simplicity. Einstein showed that Brownian particles diffuse with the diffusivity
\begin{equation}\label{EisteinD}
\k={k_BT\over 6\pi\eta R},
\end{equation}
where $R$ is the radius of the Brownian particle and $\eta$ is the viscosity of environment fluid. (Note that the viscosity $\eta$ is also a function of temperature and it is usually assumed that $\eta\propto T^{s}$ with ${1\over2}<s<1$.) Then, from the relations in (\ref{TDC}) and (\ref{SC}), the thermal diffusivity $\DT$ and the Soret coefficient $\ST$ are computed by
\begin{equation}\label{DTST}
\DT={D\over S}{dS\over dT}={D\over \sqrt{T}}{1\over2}{1\over \sqrt{T}}={k_B\over 12\pi R}\,{\eta^{-1}},\quad \ST={1\over2T}.
\end{equation}
Remember that we are considering the thermal diffusion given by random Brownian displacements only. Pollen grains in water with a temperature gradient is a good example. This non-uniform random walk analysis is the foundation of the thermal diffusion phenomenon in heterogeneous environment as a uniform one did for a homogeneous case.

\section*{Comparison to other models}
There have been many discussions and debates to find the correct diffusion flux in heterogeneous environments and the thermal diffusion phenomenon was the motivation. Einstein's relation for homogeneous case is written in two ways,
$$
\f=-{|\Delta x|^2\over 2n\Delta t}\nabla u\quad\mbox{or}\quad \f=-{k_BT\over 6\pi\eta R}\nabla u,
$$
where the first one is from the random walk point of view and the second one is from the thermodynamics point of view. Our derivation is from the first one and, however, one can find that most of other models are based on the thermodynamics point view using the kinetic theory.

The separation phenomenon of thermal diffusion was found by Ludwig \cite{Ludwig1856} for the first time and then independently by Soret \cite{Soret1879}.  Similar separation phenomena in gaseous fluid is theoretically predicted by Enskog \cite{Enskog1912} and Chapman \cite{Chapman1916} independently and then confirmed experimentally later. However, Fick's diffusion flux in (\ref{FickLaw}) does not explain the phenomenon since the only possible zero flux distribution is a constant state. There have been many efforts to extend Einstein's homogeneous diffusion theory to a heterogeneous one to explain the thermal diffusion phenomenon. For a gaseous case Chapman \cite{{Chapman1928},{MR0258399}} suggested a Fokker-Planck type diffusion flux
\begin{equation}\label{Chapman}
\f=-\nabla (\k u).
\end{equation}
In a gaseous state, Chapman's theory is widely accepted. If the flux is given by this relation, the steady state should be inversely proportional to the diffusivity. However, we have already observed in Figure \ref{fig2} that, even if the diffusivity $\k$ is constant, the steady state of a random walk system is not necessarily a constant and hence we can immediately say that Chapman's theory fails.

For liquid case, Kramers' kinetic equation \cite{Kramers1940} for the motion of Brownian particle obtained considerable attentions. For example, van Kampen \cite{KAMPEN1988} employed the equation and derived a diffusion flux $\f=-{D\over T}\nabla(Tu)$. However, temperature $T$ is proportional to the square of the particle speed, i.e., $T\propto S^2$, and hence it is against our model. It is widely accepted that there is ``no universal answer" for the diffusion flux in these kinetic theory approaches (see \cite{KAMPEN1988} for more discussions). However, the diffusion model (\ref{RW2}) is the only possible one we may obtain from a non-uniform random walk system. We could easily confirm our model by Monte Carlo simulations in the previous section.

There can be many different dynamics that are involved in the thermal diffusion. If the environment is homogeneous, their effects are canceled out and the random walk or Brownian motion is just enough to explain the diffusion. For a heterogeneous situation, these dynamics may make a difference and the situation may become complicate. However, the non-uniform Brownian displacements is the source of thermal diffusion and its theoretical foundation is still Einstein's random walk equipped with non-constant walk length and traveling time.

\section*{Conclusions}

In this paper we have shown that it is still Einstein's random walk that provides the theoretical foundation of thermal diffusion. This thermal diffusion theory is as simple as the original homogeneous one. The only thing one should add is the heterogeneity of a random walk under a temperature gradient. The diffusion flux of a random walk system has been obtained as
$$
\f=-{D\over S}\nabla(Su),
$$
where $D={|\Delta x|^2\over 2n\Delta t}$ is the diffusivity and $S={\Delta x\over\Delta t}$ is the walk speed. In thermodynamics $S$ corresponds to the instantaneous Brownian particle speed. This relation shows that the steady state of a diffusion process is not decided by the diffusivity, but by the particle speed.

The analysis of thermal diffusion in this paper is based on a random Brownian displacements which is not necessarily of a constant speed due to a temperature gradient. The classical example of pollen grains suspended in water with non-constant temperature is the case one may think of. This non-isothermal diffusion theory will provide the foundation of more general cases such as a one with two competing species. It is quite surprising that Einstein's idea still solves 156 years old Ludwig's thermophoresis. All we have to do is simply adding the heterogeneity to the walk length and jumping time, but not to the diffusivity. ``From our more distant perspective, it is clear that the Brownian-motion papers of 1905 had just as much influence on science as did relativity or light quanta. Brownian motion was just a slower, subtler revolution: not a headlong charge, but more of a random walk into a vast and unsuspected future (quoted from \cite{Haw2005})."

The random walk is being used in various fields and the theory is now equipped with non-uniform structures to handle heterogeneous environment. In fact, the non-uniform random walk has been applied to ecology models \cite{Cho12,KimKwonLi2012,KimKwonLi2013}. If a diffusion process evolves in an heterogeneous environment, this non-uniform random walk will be useful.

\providecommand{\bysame}{\leavevmode\hbox to3em{\hrulefill}\thinspace}
\providecommand{\MR}{\relax\ifhmode\unskip\space\fi MR }
\providecommand{\MRhref}[2]{%
  \href{http://www.ams.org/mathscinet-getitem?mr=#1}{#2}
}
\providecommand{\href}[2]{#2}

\end{document}